\documentstyle[preprint,amssymb,epsfig,aps]{revtex}

\begin{document}

\rightline{UCOFIS 3/00}
\rightline{US-FT/21-00}
\rightline{December 2000}
\tightenlines

\begin{center}
{\Large {\bf Baryon and strangeness enhancement at SPS, RHIC}}

{\Large {\bf and LHC energies in the String Fusion Model}}
\end{center} 

\vskip 0.6cm

\centerline{ N. S. Amelin$^{\mbox{\scriptsize{a}}}$, 
N. Armesto$^{\mbox{\scriptsize{b}}}$, 
C. Pajares$^{\mbox{\scriptsize{c}}}$ and D. Sousa$^{\mbox{\scriptsize{c}}}$}
\vskip 0.6cm

\centerline{\it $^{\mbox{\scriptsize{a}}}$ Department of Physics, University of Jyv\"askyl\"a,
FIN-40351 Jyv\"askyl\"a, Finland}

\vskip 0.3cm

\centerline{\it $^{\mbox{\scriptsize{b}}}$ Departamento de F\'{\i}sica, M\'odulo C2, Planta Baja,
Campus de Rabanales,} 
\centerline{\it Universidad de C\'ordoba, E-14071 C\'ordoba, Spain}

\vskip 0.3cm

\centerline{\it $^{\mbox{\scriptsize{c}}}$ Departamento de F\'{\i}sica de Part\'{\i}culas
Elementales,}
\centerline{\it Universidade de Santiago de Compostela, E-15706 Santiago de Compostela, Spain}

\vskip 0.6cm

\begin{abstract}
Strangeness and baryon enhancement in heavy ion collisions are discussed in
the framework of the String Fusion Model. The Monte Carlo version of this model is
shown to reasonably reproduce three of the features that have been pointed out
as evidences of the finding of the Quark Gluon Plasma. Namely hyperon/antihyperon
enhancement, dependence of the slope of the $p_{T}$ distributions on the mass of the
produced particles and on the centrality of the collision, are described. Predictions
for RHIC and LHC energies are presented.
\end{abstract}

\newpage
\section{Introduction and model description}

In the last year a large excitement has arisen in the 
heavy ion physics community, related to
the possibility of Quark Gluon Plasma (QGP) 
already been obtained
at SPS energies \cite{cern}.
In particular several signals were mentioned, which point out to the existence of
QGP. Putting aside the abnormal $J/\psi$ suppression and the excess of dileptons
found, there are three signals related to baryon and strangeness production, namely the
large enhancement of the (anti)hyperon yields ($\Lambda$, $\Xi$, $\Omega$) in Pb-Pb
collisions compared to p-Pb, observed by the WA97 \cite{wa97} and the NA49 \cite{na49} 
Collaborations\footnote{A recent reanalysis \cite{reanalysis} of $\Xi$
data done by the NA49 Collaboration gives yields at midrapidity which are
in much closer agreement to the WA97 \cite{wa97} results than the
previous analysis of NA49 \cite{na49}.};
the linear increase of the inverse exponential
slope of the $p_{T}$ distributions ('temperature') in Pb-Pb
collisions with the mass of the observed particle, except
for $\Omega$ (for pions there is also a small departure from the straight line)
\cite{na49,pt}; 
and the different behaviour of the temperature between
p-p and A-A collisions. 
Concerning this last feature,
while the temperature
in p-p collisions remains flat as a function of particle mass,
in heavy ion collisions it increases as the mass increases.
Furthermore,
for a given mass, the heavier the colliding system, the higher the temperature \cite{na44}.
These last two characteristics have been interpreted as the existence of an intrinsic
freeze-out temperature and a collective hydrodynamical flow which is gradually developed:
firstly, for S-S collisions, and, in a more clear way, in Pb-Pb collisions.

We would like to study these three signals, together with other important related data as
$\phi$ production \cite{phi1,phi2}, 
different particle ratios \cite{ratios} and stopping power \cite{stop}, in a Monte Carlo code based 
on the String Fusion Model (SFM) \cite{sfm1,sfm2}. 
Compared to previous versions, the model has been improved, including minijet production
and rescattering of the produced secondaries in a standard way (an extensive description
of these new features will be presented elsewhere \cite{code}). Summarizing, hard
collisions have been included using PYTHIA \cite{pythia}, in order 
to reproduce the non-exponential
tail of the $p_{T}$ spectrum experimentally
measured in \=p-p collisions at Sp\=pS and TeVatron; this is crucial
for the applicability of the model at RHIC and LHC. Rescattering is introduced as 
$2 \rightleftarrows 2$ collisions \cite{rescat} between mesons and baryons, 
taking into account both strangeness
conserving and strangeness exchange reactions, and the possibility of the inverse
processes as required by detailed balance. Only three cross sections are used:
the first one for reactions involving $\Omega$ and $\bar\Omega$, the second one for
reactions related to $\bar p$ annihilation, and the third one for all other reactions.

In the
String Fusion Model it is assumed that strings fuse when their transverse positions
come within a certain interaction area. In the code we consider
only fusion of two strings but there is a probability of fusion of more than two. An
effective way of taking this into account is to increase the cross section for the fusion of
two strings, for which  we will take $\sigma_{fus} = 7.5$ mb. This value is crucial to 
reproduce the strangeness enhancement 
in central S-S and S-Ag collisions
at SPS \cite{sfm2}. The fusion can take 
place only when the rapidity intervals of the strings overlap. It is formally described by
allowing partons to interact several times, the number of interactions being the same 
both for projectile and target. The quantum numbers of the fused strings are determined by the
interacting partons and their energy-momentum is the sum of the energy-momenta of the
ancestor strings. The colour charges of the fusing string ends sum into the colour charge
of the resulting string ends according to the SU(3) composition laws. The breaking of
each fused string is due to the production of two (anti)quark complexes with the same 
colour charges $Q$ and $\bar Q$ as those at the 
ends of the string \cite{sfm1}. The probability
rate is given by the Schwinger formula \cite{schwin}
\begin{equation}
W \sim K^{2}_{\{N\}} \textnormal{exp}(-\pi M_{t}^{2} / K_{\{N\}}),
\end{equation}
where $K_{\{N\}}$ is the string tension for the $\{N\}$ SU(3) representation,
proportional to the corresponding quadratic Casimir operator $C^{2}_{\{N\}}$ \cite{bali}. In our
case $C^{2}_{\{3\}} = 4/3$, $C^{2}_{\{6\}} = 10/3$ and $C^{2}_{\{8\}} = 3$.
Therefore, the $\{8\}$ and $\{6\}$ fused strings have a higher string tension,
giving rise to larger heavy flavour and baryon/antibaryon production.
Hard strings are not fused, their area being proportional to
$1/p_{T}^{2}$. Some effect of the fusion of such strings could appear
at LHC energies where, for instance, in central Pb-Pb collisions they amount for 35 \% of the 
binary nucleon-nucleon collisions \cite{code}. 

In principle, the fusion of strings means nothing
related to a phase transition. On the contrary, percolation of strings \cite{perc} is
a non-thermal second order phase transition. In this case, the key parameter is
$\eta = \pi r^{2} N / A$, which is the density of strings $N/\pi R_A^{2}$ (number of
strings $N$ produced in the overlapping area of the collision, $A = \pi R_A^{2}$
for central collisions) times the
transverse size of one string $\pi r^{2}$. The critical point 
for percolation is $\eta_{c} \simeq
1.12 \div 1.5$ depending on the profile function of 
the colliding nuclei \cite{dias}. With $r \simeq 0.2$ fm,
this critical value means $9 \div 12$ strings/fm$^{2}$. The value of 9 is reached 
in central Pb-Pb collisions at SPS, in central Ag-Ag collisions at RHIC and in central
S-S collisions at LHC. We expect for $\eta$ around or greater than $\eta_{c}$, that
the approximation of fusion of just two strings fails. 

\section{Results and conclusions}

In Fig. 1 we show our results for $\Omega$, $\Xi$ and
$\Lambda$ yields for p-Pb,
and central Pb-Pb collisions at SPS with four different centralities,
together with the experimental
data. In order to disentangle the different processes contributing, in 
Fig. 2 it is shown the results of the code 
for central ($b \le 3.2$ fm) Pb-Pb collisions without
string fusion and rescattering, with string fusion, and with string fusion and rescattering.
A reasonable agreement with data is obtained. 
Only the $\Omega$'s are a little 
below the data (40 \%).
Similar results have been obtained in the Relativistic Quantum Molecular Dynamics
model \cite{rqmd,last} by a mechanism of colour ropes which consider fusion of
strings; also in the Ultra Relativistic Quantum Molecular
Dynamics model \cite{rqmd1,rqmd2}
and in the HIJING model \cite{vance}
by using an ad hoc multiplicative factor in the string 
tension. Also
the Dual Parton Model \cite{dpm}, considering the possibility of creation of 
diquark-antidiquark pairs in the nucleon sea,
together with the inclusion of diagrams which take into account baryon 
junction migration \cite{dpm,bj1,bj2}, can reproduce the experimental 
data (for $\Omega$'s some rescattering has still to be added).
The string fusion is the main ingredient to obtain a 
good agreement with $\bar \Lambda$ experimental data and also to reproduce
the $\Xi$ data. However the rescattering is fundamental to get enough $\Omega$'s. Notice also
that there is not any jumping between S-S and Pb-Pb. The most pronounced enhancement
takes place between p-Pb and S-S. About $\bar \Lambda$, our results are higher than
the WA97 data, its production being mainly determined by string fusion and hardly
affected by rescattering. This fact makes that our results for Pb-Pb are really an
extrapolation in the model from the value for $\bar \Lambda$ production in
central S-S collisions by the NA35 Collaboration, which was used to fix the fusion 
cross section $\sigma_{fus}$ \cite{sfm2}. So, from the point of view of our model,
there exists either a large $\bar \Lambda$ annihilation or a conflict between
NA35 data for S-S and WA97 data for Pb-Pb.

In Fig. 3 we plot the inverse exponential slopes of the $p_{T}$ 
distributions for different particles, together with the
WA97 experimental data.
A reasonable agreement is obtained. In particular it can be seen that the
$\Omega$ slope is below the straight line both in the model and in data. 
Also our predictions
for RHIC are presented. We see that the $\Omega$ is now on the straigth line,
so the decrease at SPS energies is due to energy-momentum conservation. 

About $\phi$ enhancement, our results without fusion, with fusion, and
with fusion and
rescattering are 3.55, 4.20 and 5.35 respectively, in satisfactory agreement with
experimental data, $7.6 \pm 1.1$ \cite{phi2}.
In Fig. 4 the stopping power is shown, i.e. the $p-\bar p$ rapidity distributions for
central Pb-Pb collisions at SPS, compared with 
the experimental data \cite{stop}, together with the predictions
for RHIC and LHC energies. This quantity is essentially determined by the string fusion
mechanism and rescattering only plays a minor role. As discussed for strangeness enhancement,
it has been pointed out
that baryon junction migration \cite{dpm,bj1,bj2} will enhance the stopping 
power due to
diagrams additional to the usual ones of the Dual Parton Model.
The inclusion of these diagrams also explains the SPS data. We have not 
taken into account such diagrams to avoid double counting, because in the
fusion of strings they are partially included in an effective way.
In Fig. 5 the antiproton rapidity distribution in central Pb-Pb
collisions is presented and compared to the experimental data \cite{aprot}; a great
suppression of the antiproton yield is seen, due to rescattering.

In Table I our results for the ratios between different particles are compared
with the experimental data \cite{ratios} for Pb-Pb central collisions at SPS. 
We include our predictions for
RHIC and LHC in Table II. Some comments are in order:
First, we observe an overall, rough
agreement  with the SPS data. Second,
our results are not very 
different to those of statistical models \cite{stat1,stat2}. However, the saturation
of the strangeness enhancement in our case has nothing to 
do with thermal and/or chemical equilibrium.
In string fusion the enhancement of the different strange particle yields is similar 
to a threshold behaviour: First there is a pronounced rise and afterwards, a saturation. The 
main difference in the predictions for RHIC and LHC between the String Fusion Model
and statistical models
is the overall charged multiplicity, which is respectively 
950 and 3100 for SFM and 1500 and 7600 for
statistical models \cite{rev} (assuming initial temperatures of 500 and
1000 MeV for RHIC and LHC respectively).

Let us emphasize that we obtain a reasonable agreement with the experimental data in three of the
features advocated as signals of QGP production. 
We are only below data in the $\Omega$ production by less than a 
factor 2. It is possible to improve this 
agreement arranging the corresponding rescattering cross sections. However, we think that
the usefulness of any Monte Carlo is in the overall picture and not in obtaining
detailed fits which require a fine tuning. 
For this reason we do not change our rescattering cross sections.

Nevertheless, related to the $\Omega$ production, it is tempting to speculate about 
percolation of strings as a possible explanation of the discrepancy. It
is just around Pb-Pb central collisions at SPS energies where the critical point
$\eta_{c}$, which corresponds to $9 \div 12$ strings/fm$^{2}$, is placed. Close to this point,  
the formation of at least one cluster of strings through the whole 
reaction area is possible. 
Therefore it is expected that the String Fusion Model does not work there. This point
can be checked at RHIC. There, the density of strings formed in central Ag-Ag collisions
is also around 9 and therefore the number of $\Omega$'s
would be above the string
fusion prediction. 

As a last point, we compare our results with the preliminary data of the PHOBOS Collaboration
at RHIC \cite{rhic}. For charged particles we 
obtain $dN/d\eta \mid_{\mid \eta \mid < 1} = 520$ and 585
for the 6 \% more central Au-Au collisions at $\sqrt{s} = 56$ and 130 GeV per nucleon respectively,
to be compared with $408 \pm 12\ {\rm (stat)} \pm 30\ {\rm (syst)}$
and $555 \pm 12\ {\rm (stat)} \pm 35\ {\rm (syst)}$. Our prediction for
$\sqrt{s} = 200$ GeV per nucleon with the same centrality criterium is
$dN/d\eta \mid_{\mid \eta \mid < 1} = 635$.

In conclusion, the agreement of the model with baryon and strangeness 
experimental data at SPS is satisfactory, so the 
SFM code can be useful as a tool for
simulations of the forthcoming experiments at RHIC and LHC energies.   

\acknowledgements

We thank M. A. Braun and E. G. Ferreiro, who participated in early stages
of this work. We also thank A. Capella, A. B. Kaidalov, C. A. Salgado and
Yu. M. Shabelski for useful discussions, and J. Stachel for comments on
the predictions of the statistical models for RHIC and LHC.
N. A., C. P. and D. S. 
acknowledge financial support by CICYT of Spain under contract
AEN99-0589-C02 and N. S. A. by Academy of Finland under grant 
number 48477. N. A. thanks Departamento de F\'{\i}sica de
Part\'{\i}culas of the Universidade de Santiago de Compostela, and D. S.
the ALICE Collaboration at CERN, for hospitality during stays in which
part of this work was completed.

\newpage

\begin{table}[hbtp]
\begin{displaymath}
\begin{array}{||c|c|c|c|c||}
\hline
\hline
& NF & F & F+R & Exp.\\
\hline
\hline
\bar \Lambda/\Lambda & 0.036 & 0.256 & 0.202 & 0.128 \pm 0.012 \\ \hline
\bar \Xi^{+}/\Xi^{-} & 0.959 & 0.414 & 0.342 & 0.266 \pm 0.028 \\ \hline
\bar \Omega^{+}/\Omega^{-} & 1.0 & 0.543 & 0.454 & 0.46 \pm 0.15 \\ \hline
\Xi^{-}/\Lambda & 2.58 \cdot 10^{-3} & 0.042 & 0.067 & 0.093 \pm 0.007 \\ \hline
\bar \Xi^{+}/\bar \Lambda & 0.069 & 0.069 & 0.113 & 0.195 \pm .023 \\ \hline
\Omega/\Xi & 0.0606 & 0.046 & 0.106 & 0.195 \pm .028 \\
\hline
\hline
\end{array}
\end{displaymath}
\caption{SFM results for different particle ratios in
central Pb-Pb collisions at SPS, compared with experimental data
$(Exp.)$
[9].
Results are presented without fusion (NF),
with fusion (F) and with fusion and rescattering (F+R).}
\label{table2}
\end{table}

\begin{table}[hbtp]
\begin{displaymath}
\begin{array}{||c|c|c|c|c|c|c||}
\hline
\hline
& RHIC (F) & RHIC (F+R) & QCM & Rafelski & B-M & LHC (F)\\
\hline
\hline
\bar \Lambda/\Lambda & 0.638 & 0.600 & 0.84 & 0.49 \pm 0.15 & 0.906 & 0.889 \\ \hline
\bar \Xi^{+}/\Xi^{-} & 0.860 & 0.823 & 0.84 & & 1. & 0.965 \\ \hline
\bar \Omega^{+}/\Omega^{-} & 1.207 & 0.975 & 1. & 1. & 1. & 1.047 \\ \hline
\Xi^{-}/\Lambda & 0.071 & 0.108 & & & 0.125 & 0.084 \\ \hline
\bar \Xi^{+}/\bar \Lambda & 0.096 & 0.149 & & 0.18 \pm 0.2 & 0.138 & 0.091 \\ \hline
\Omega/\Xi & 0.078 & 0.208 & & & & 0.055 \\ \hline
\bar \Lambda/\bar p & 0.386 & 0.726 & 0.87 & 2.4 \pm 0.3 & & 0.355 \\ \hline
\bar p \ p & 0.499 & 0.380 & 0.67 & & 0.097 & 0.829 \\
\hline
\hline
\end{array}
\end{displaymath}
\caption{SFM results for different particle ratios in
central Pb-Pb collisions at RHIC and LHC,
following the same convention as in Table
I. For comparison, results from other models (Quark Coalescence Model (QCM)
[21],
Rafelski
[21]
 and $B-M$
[21,29])
for RHIC are 
included.}
\label{table3}
\end{table}

\newpage

\begin{figure}
\epsfxsize=10cm
\begin{center}
\mbox{\epsfig{file=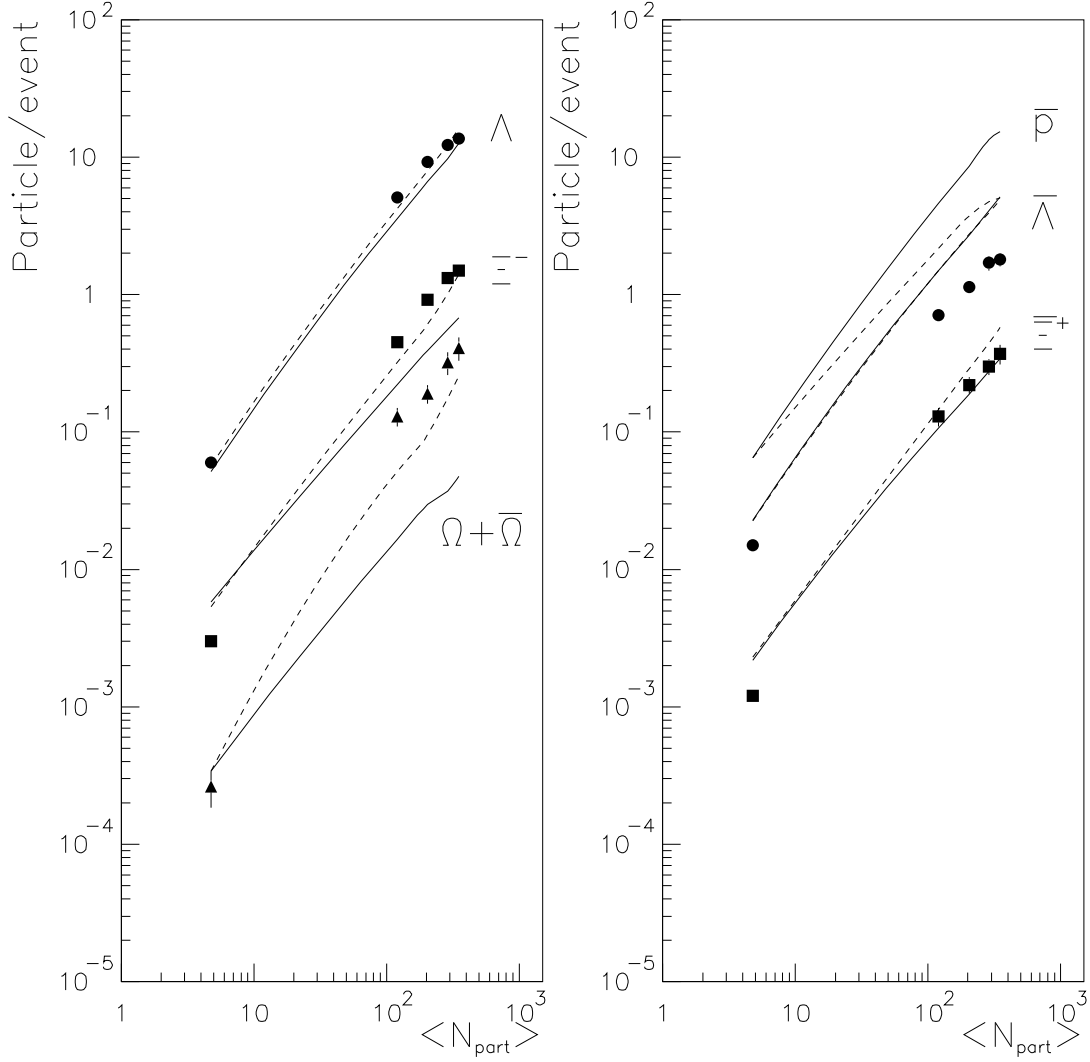,width=16cm}}
\end{center}
\caption{Yields per unity of rapidity at central rapidity as a function
of the number of wounded nucleons for $\Lambda$, $\Xi^{-}$ and $\Omega + \bar\Omega$
(left) and for $\bar p$, $\bar\Lambda$ and $\bar\Xi^{+}$ (right) for p-Pb collisions and
four different centralities in Pb-Pb collisions at SPS energies. Full lines
represent our calculation with string fusion and dashed lines with fusion and rescattering.
Experimental data are from the WA97 Collaboration
[2].
}
\label{pbpbext}
\end{figure}

\newpage

\begin{figure}
\epsfxsize=10cm
\begin{center}
\mbox{\epsfig{file=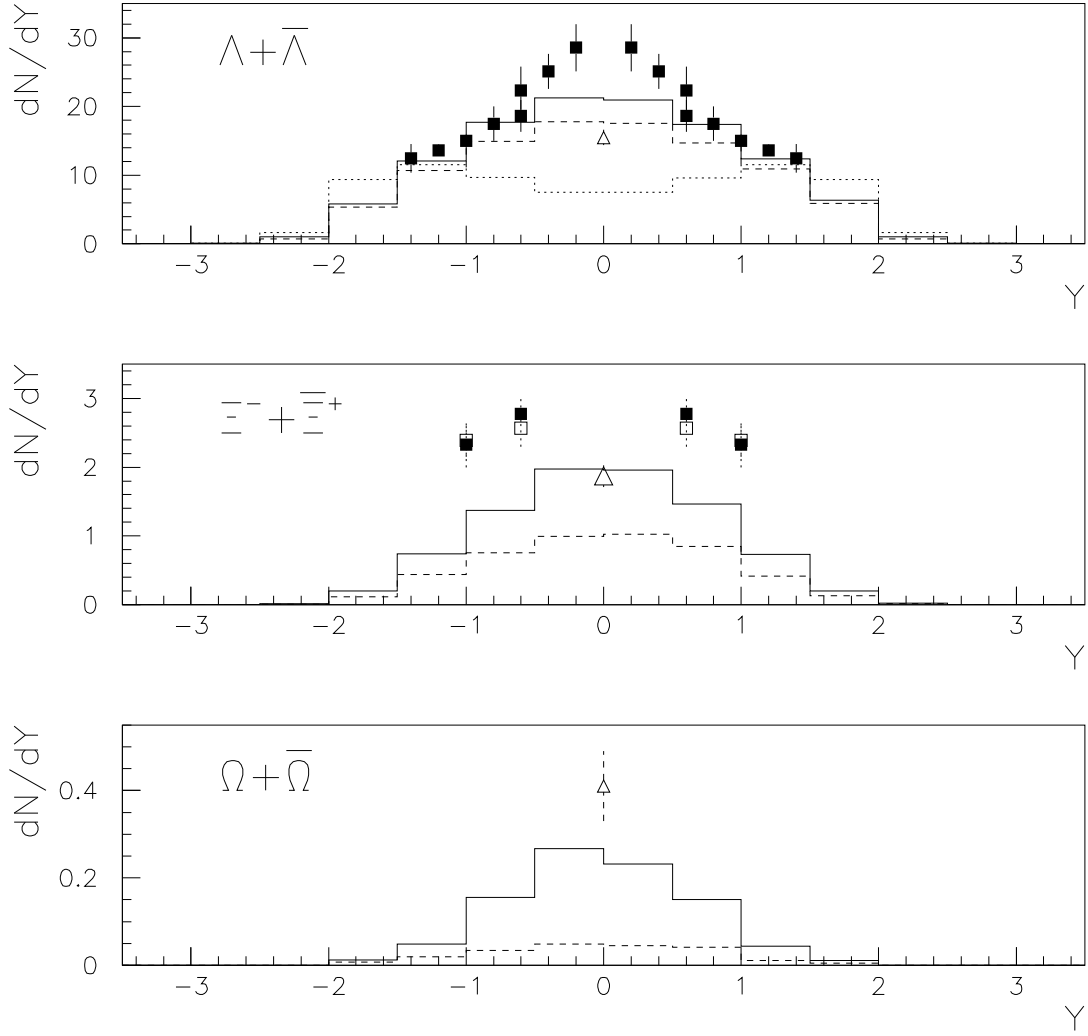,width=16cm}}
\end{center}
\caption{SFM results (dotted line: without fusion, dashed line: with
fusion, solid line: with fusion and rescattering)
for strange baryon production in central Pb-Pb
collisions (5 \% centrality) at SPS compared with experimental data
from the WA97 Collaboration
[2]
(triangles) and the NA49 Collaboration
[3]
 (squares).}
\label{extra}
\end{figure}

\newpage

\begin{figure}
\epsfxsize=10cm
\begin{center}
\mbox{\epsfig{file=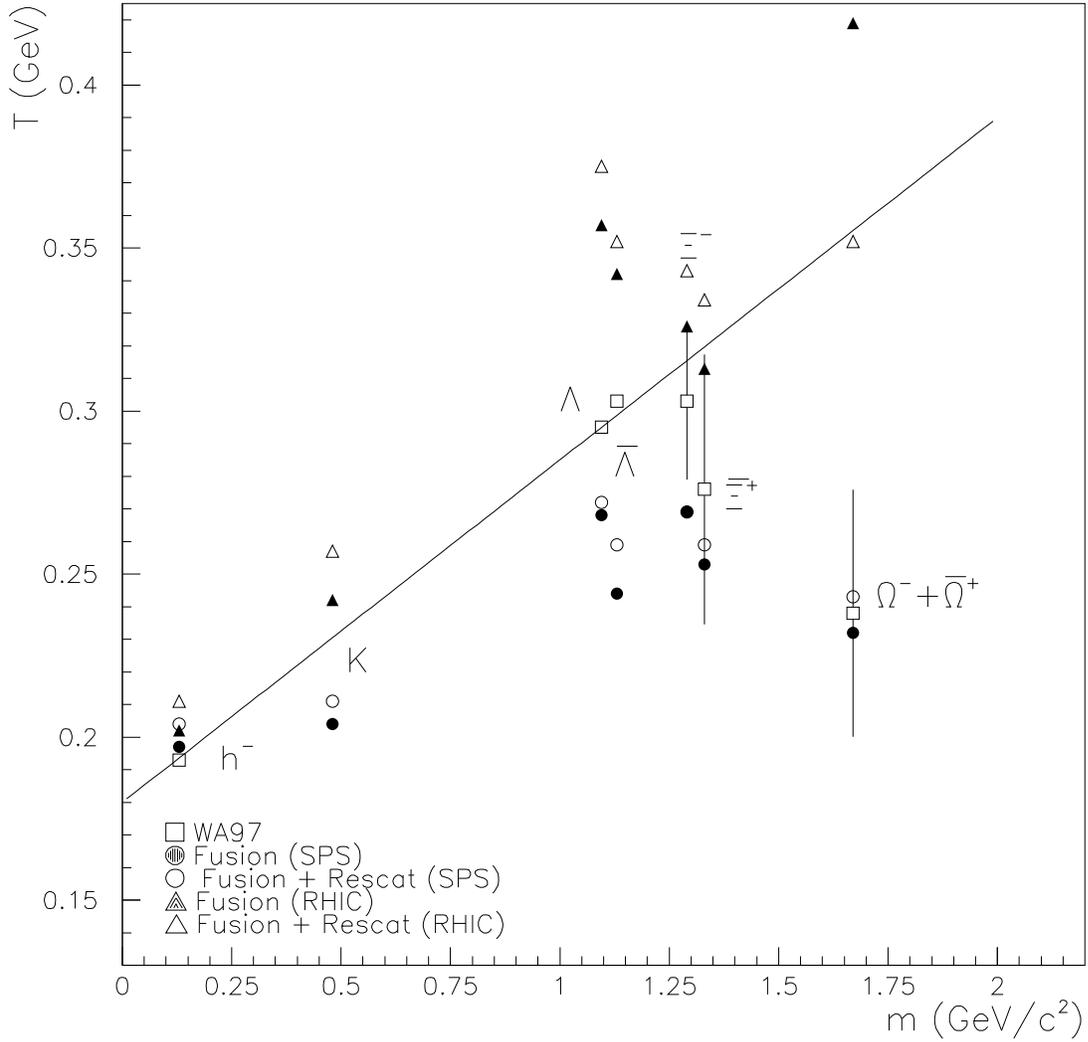,width=16cm}}
\end{center}
\caption{SFM results (filled circles: with fusion, open circles: with
fusion and rescattering) for the inverse exponential
slope of the $p_{T}$ distributions of different particles
versus the transverse mass of the
particles in central (5 \% centrality) Pb-Pb collisions at SPS,
compared with the experimental data of the WA97 Collaboration
[5]
(open squares). We also present our predictions for
RHIC energy with fusion, filled triangles, and with fusion and rescattering,
open triangles.}
\label{temp}
\end{figure}

\newpage

\begin{figure}
\epsfxsize=10cm
\begin{center}
\mbox{\epsfig{file=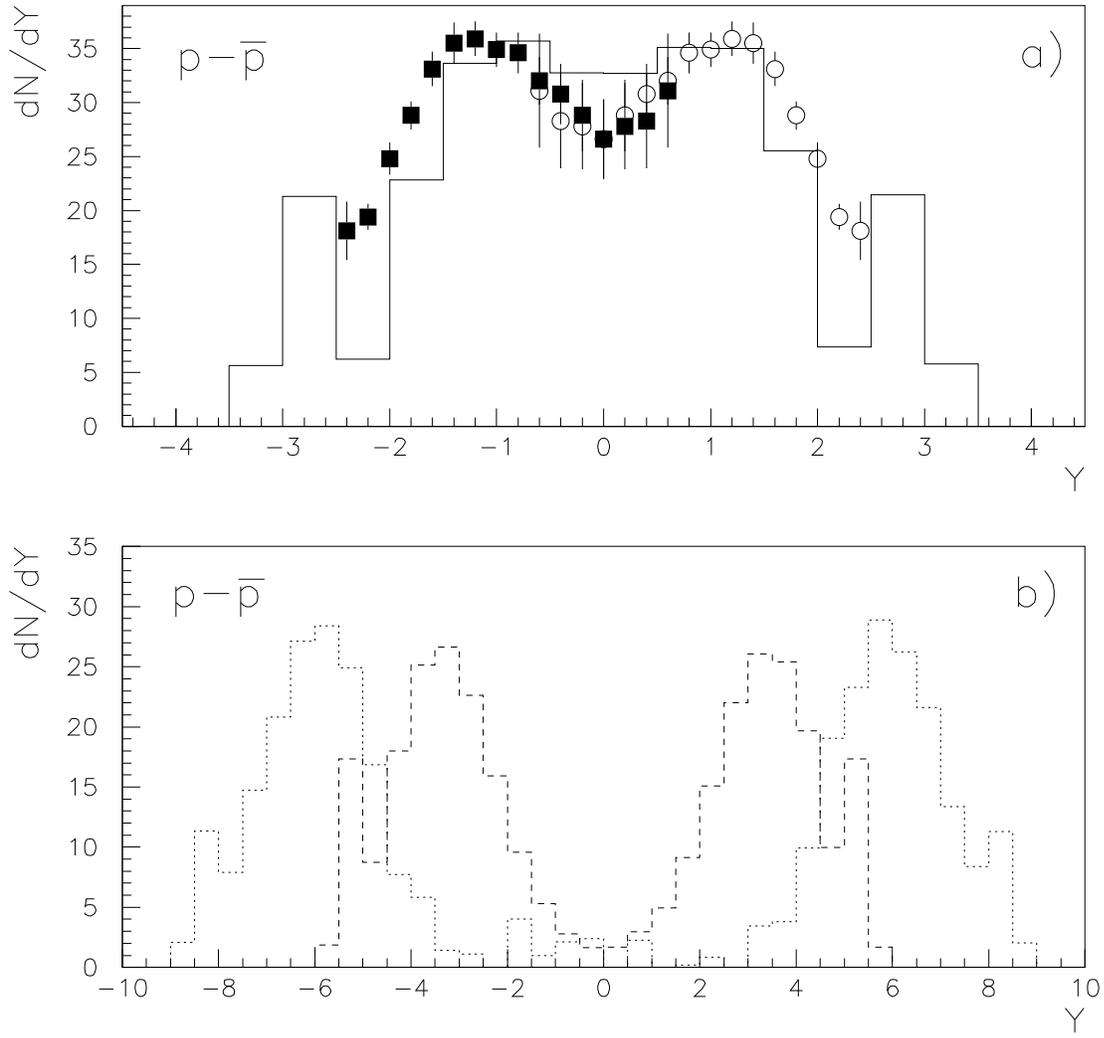,width=16cm}}
\end{center}
\caption{SFM results for the $p-\bar p$ rapidity distribution in central
(5 \% centrality) Pb-Pb collisions at SPS (solid line), RHIC (dashed
line) and LHC (dotted line), compared with experimental data at SPS
[10].
}
\label{stop}
\end{figure}

\newpage

\begin{figure}
\epsfxsize=10cm
\begin{center}
\mbox{\epsfig{file=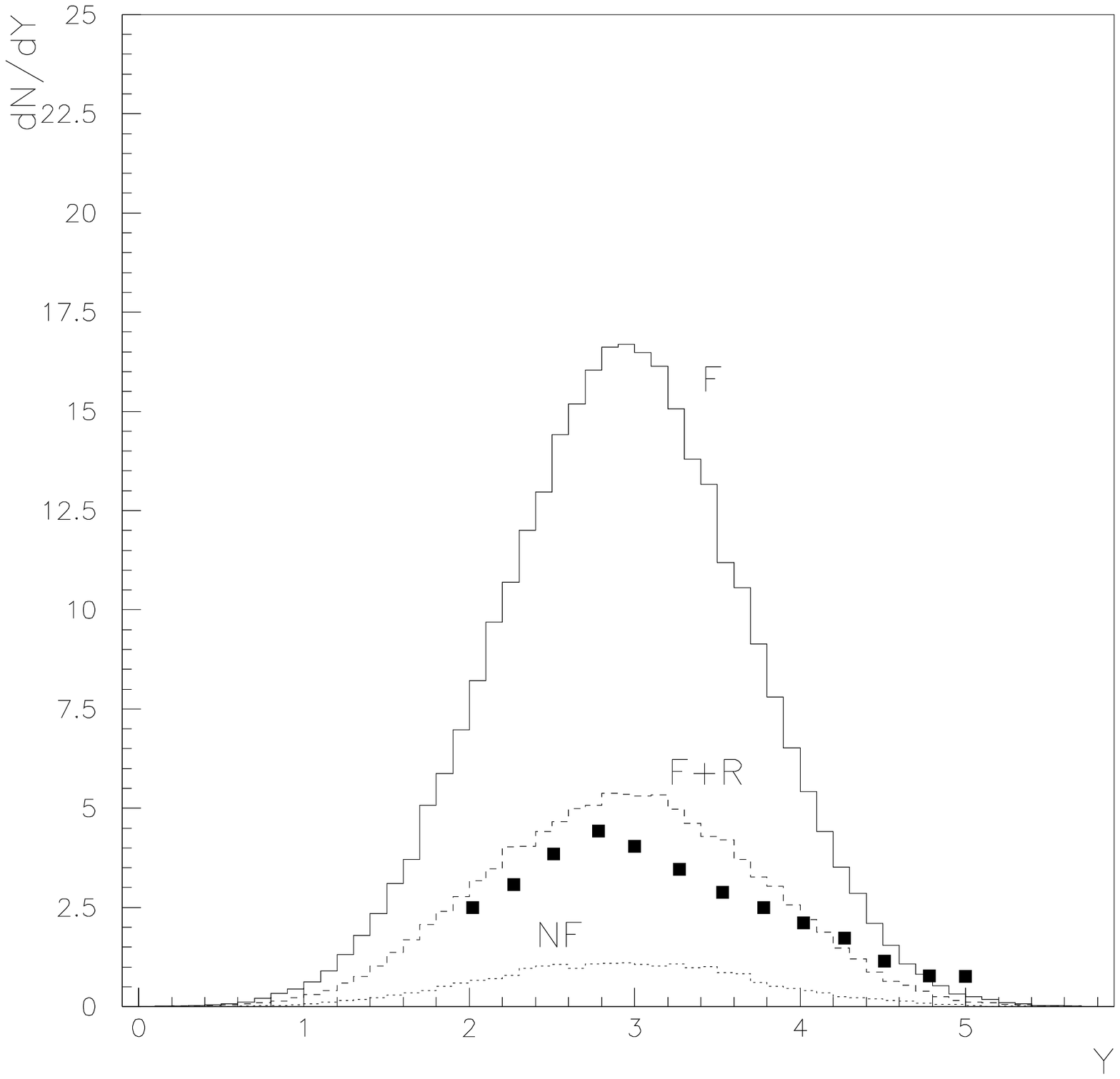,width=16cm}}
\end{center}
\caption{SFM results (dotted line: without fusion, 
solid line: with fusion, dashed line: with fusion
and rescattering) for the $\bar p$ rapidity distribution in central
(5 \% centrality) Pb-Pb collisions at SPS, compared with experimental
data
[28].
}
\label{aprot}
\end{figure}


\begin{thebibliography}{99}

%
\bibitem{cern} http://www.cern.ch/CERN/Announcements/2000/NewStateMatter/
%
\bibitem{wa97} WA97 Collaboration: E. Andersen {\it et al.}, Phys. Lett. {\bf B433}
(1998) 209; {\it ibid.} {\bf B449} (1999) 401.
%
\bibitem{na49} NA49 Collaboration: H. Appelsh\"auser {\it et al.}, Phys. Lett. {\bf B444}
(1998) 523; Eur. Phys. J. {\bf C2} (1998) 661. 
%
\bibitem{reanalysis} NA49 Collaboration: R. A. Barton {\it et al.}, in
{\it Proceedings of the Strangeness 2000 Conference} (Berkeley, USA, July 
20th-25th 2000).
%
\bibitem{pt} WA97 Collaboration: F. Antinori {\it et al.}, Nucl. Phys. 
{\bf A661} (1999) 481c; Eur. Phys. J. {\bf C14} (2000) 633. 
%
\bibitem{na44} NA44 Collaboration: N. Xu {\it et al.}, Nucl. Phys. {\bf A610}
(1996) 175c. 
%
\bibitem{phi1} NA50 Collaboration: N. Willis {\it et al.}, 
Nucl. Phys. {\bf A661} (1999) 534c. 
%
\bibitem{phi2} NA49 Collaboration: G. H\"ohne {\it et al.}, 
Nucl. Phys. {\bf A661} (1999) 485c; S. V. Afanasev {\it et al.},
Phys. Lett. {\bf B491} (2000) 59.
%
\bibitem{ratios} WA97 Collaboration: I. Kr\'alik {\it et al.},
Nucl. Phys. {\bf A638} (1998) 115c.
%
\bibitem{stop} NA49 Collaboration: G. E. Cooper {\it et al.}, 
Nucl. Phys. {\bf A661} (1999) 362c. 
%
\bibitem{sfm1} N. S. Amelin, M. A. Braun and C. Pajares, Phys. Lett. {\bf B306}
(1993) 312; Z. Phys. {\bf C63} (1994) 507. 
%
\bibitem{sfm2} N. Armesto, M. A. Braun, E. G. Ferreiro and C. Pajares,
Phys. Lett. {\bf B344} (1995) 301; E. G. Ferreiro, C. Pajares and D. Sousa,
Phys. Lett. {\bf B422} (1998) 314. 
%
\bibitem{code} N. S. Amelin, N. Armesto, C. Pajares and D. Sousa,
in preparation. 
%
\bibitem{pythia} T. Sj\"ostrand, Comput. Physics Commun. {\bf 82} (1994) 74.
%
\bibitem{rescat} P. Koch, B. M\"uller and J. Rafelski, Phys. Rep. {\bf 142}
(1986) 167.
%
\bibitem{schwin} J. Schwinger, Phys. Rev. {\bf 82} (1951) 664; E. Brezin and
C. Itzykson, Phys. Rev. {\bf D2} (1970) 1191.
%
\bibitem{bali} G. S. Bali, Nucl. Phys. Proc. Suppl. {\bf 83} (2000) 831;
preprint HUB-EP-99-67 (hep-ph/0001312).
%
\bibitem{perc} N. Armesto, M. A. Braun, E. G. Ferreiro and C. Pajares,
Phys. Rev. Lett. {\bf 77} (1996) 3736. 
%
\bibitem{dias} A. Rodrigues, R. Ugoccioni and J. Dias de Deus,
Phys. Lett. {\bf B458} (1999) 402.
%
\bibitem{rqmd} H. Sorge, Phys. Rev. {\bf C52} (1995) 3291.
%
\bibitem{last} S. A. Bass {\it et al.}, Nucl. Phys. {\bf A661} (1999) 205c.
%
\bibitem{rqmd1} M. Bleicher, M. Belkacem, S. A. Bass, S. Soff and H. St\"ocker,
Phys. Lett. {\bf B485} (2000) 213.
%
\bibitem{rqmd2} M. Bleicher, W. Greiner, H. St\"ocker and N. Xu, hep-ph/0007215. 
%
\bibitem{vance} S. E. Vance, in {\it Proceedings of the Strangeness 2000
Conference} (Berkeley, USA, July 20th-25th 2000), nucl-th/0012056.
%
\bibitem{dpm} A. Capella and C. A. Salgado, Phys. Rev. {\bf C60} (1999) 054906;
preprint LPT-ORSAY-00-66 (hep-ph/0007236); 
A. Capella, E. G. Ferreiro and C. A. Salgado,
Phys. Lett. {\bf B459} (1999) 27. 
%
\bibitem{bj1} S. E. Vance and M. Gyulassy, Phys. Rev. Lett. {\bf 83} (1999) 1735. 
%
\bibitem{bj2} A. Capella and B. Z. Kopeliovich, Phys. Lett. {\bf B381} (1996) 325;
B. Z. Kopeliovich and B. Povh, Phys. Lett. {\bf B446} (1999) 321;
D. Kharzeev, Phys. Lett. {\bf B378} (1996) 238; F. W. Bopp, hep-ph/0002190. 
%
\bibitem{aprot} NA49 Collaboration: J. Bachler {\it et al.},
Nucl. Phys. {\bf A661} (1999) 45c.
%
\bibitem{bm} P. Braun-Munzinger, I. Heppe and J. Stachel, Phys. Lett. {\bf B465}
(1999) 15.
%
\bibitem{stat1} P. Braun-Munzinger, Nucl. Phys. {\bf A661} (1999) 261c. 
%
\bibitem{stat2} J. Stachel, in {\it Proceedings of the XXIXth International 
Symposium on Multiparticle Dynamics}
(Providence, USA, August 9th-13th 1999), to be published by World Scientific. 
%
\bibitem{rev} N. Armesto and C. Pajares, Int. J. Mod. Phys. {\bf A15} (2000) 2019.
%
\bibitem{rhic} PHOBOS Collaboration: B. B. Back {\it et al.},
Phys. Rev. Lett. {\bf 85} (2000) 3100.
%
\end{thebibliography}
\end{document}